# A Lightweight GAN-Based Image Fusion Algorithm for Visible and Infrared Images

1st Zhizhong Wu
University of California, Berkeley
Berkeley, USA

2nd Jiajing Chen
New York University
New York, USA

3rd LiangHao Tan
Independent Researcher
San Jose, USA

4th Hao Gong*
Independent Researcher
Shanghai, China

5th Zhou Yuru
Southwest University of Science and Technology
Mianyang, China

6th Ge Shi
Independent Researcher
San Jose, USA

*Abstract*— This paper presents a lightweight image fusion algorithm specifically designed for merging visible light and infrared images, with an emphasis on balancing performance and efficiency. The proposed method enhances the generator in a Generative Adversarial Network (GAN) by integrating the Convolutional Block Attention Module (CBAM) to improve feature focus and utilizing Depthwise Separable Convolution (DSConv) for more efficient computations. These innovations significantly reduce the model's computational cost, including the number of parameters and inference latency, while maintaining or even enhancing the quality of the fused images. Comparative experiments using the M3FD dataset demonstrate that the proposed algorithm not only outperforms similar image fusion methods in terms of fusion quality but also offers a more resource-efficient solution suitable for deployment on embedded devices. The effectiveness of the lightweight design is validated through extensive ablation studies, confirming its potential for real-time applications in complex environments.

## I. INTRODUCTION

Multi-modal image fusion aims to integrate complementary information from different sensors to generate an image that combines features of different scales and focal lengths to more comprehensively characterize target scene information, providing solutions such as target detection [1], semantic segmentation [2] and It lays a solid foundation for computer vision applications such as scene perception [3]. The fusion of visible image (VI) and infrared image (IR) is a widely used multi-modal image fusion task. Its goal is to create an image that contains both rich texture information in visible images and infrared images[4]. For YOLO small target detection [5], which is a common problem of missed detection and false detection in industrial applications [6], the use of fused images can improve detection accuracy and simplify the deployment process. Commonly used image fusion frameworks include frameworks based on Autoencoder (AE) [7], Convolutional Neural Networks (CNN) [8], and Generative Adversarial Network (GAN) [9]. In particular, GAN-based frameworks have attracted much attention due to their ability to effectively model data distribution without supervision.

In the evolving landscape of multi-modal image fusion using generative adversarial networks (GANs), the FusionGAN framework has been a pivotal development, introducing the concept of transforming the fusion challenge of visible and infrared images into a dynamic confrontation between the generator and the discriminator for the first time, as documented in references [10-11]. However, while FusionGAN was groundbreaking, it encountered notable challenges, particularly with information loss in its intermediate and shallow layers. This issue tends to exacerbate as additional layers are incorporated into the network, further amplifying the model's volume and complicating its architecture.

To address these shortcomings, the research community has embarked on refining the GAN models by incorporating a variety of advanced loss functions and making strategic enhancements to the network structure. Significant among these advancements are the introduction of the Patch-GAN model and the multi-class generative adversarial network. These models, along with structural enhancements such as the implementation of guided filters, have markedly improved the quality of the fused images. Nonetheless, these improvements often come with the trade-off of increased model size, which poses substantial deployment challenges and extends computation times, particularly affecting the practical deployment in real-time applications. To mitigate these issues and enhance the practical utility of GAN-based image fusion, especially in real-time settings, this paper introduces the Target-Aware Dual Adversarial Learning (TADAL) network. This network is specifically engineered to ensure efficient fusion of visible light images in complex environments while maintaining a lightweight structure suitable for deployment on embedded devices, thus preserving high detection performance. The core innovation in the TADAL architecture involves the integration of the Convolutional Block Attention Module (CBAM) and the replacement of standard convolution layers with Depthwise Separable Convolution (DSConv). These modifications not only reduce the computational load but also decrease the parameter count significantly.

The effectiveness of these architectural enhancements has been rigorously tested through a series of ablation experiments. The outcomes confirm that the adaptations significantly enhance the performance of the image fusion process without burdening the system resources. Further, comparative analysis with four similar image fusion algorithms demonstrates that the TADAL network offers superior fusion image quality while operating with fewer parameters compared to its counterparts. This balance of efficiency and performance underscores the potential of lightweight GAN models in overcoming the traditional barriers of large-scale deployment and lengthy computation times, setting a new standard for future developments in the field of image fusion technology.

## II. RELATED WORK

In Generative Adversarial Networks (GANs) have become a cornerstone in the development of advanced image processing techniques, particularly in multi-modal image fusion tasks. This section reviews key works that have contributed to the field and influenced the development of the proposed lightweight GAN-based image fusion algorithm. The comparative study by Zhong et al. [12] highlights the superior performance of GANs over traditional image recognition algorithms. Their work underscores the robustness of GANs in handling complex data distributions, which is fundamental to the image fusion tasks addressed in this paper. In the domain of medical imaging, Feng et al. [13] demonstrated how GANs could effectively synthesize realistic images from limited data, a technique that has proven valuable in enhancing image quality. Similarly, Liu et al. [14] explored the application of multi-modal fusion techniques in disease recognition, showcasing how GANs can integrate diverse data sources to improve diagnostic accuracy. The research by Hu et al. [15] delves into multi-scale image fusion systems, particularly in medical image analysis. Their findings emphasize the importance of maintaining fine details while minimizing computational load, a challenge that is directly addressed by the proposed lightweight modifications in the GAN architecture. In terms of addressing information loss in deep learning models, Gao et al. [16] proposed enhancements to encoder-decoder architectures to retain critical information during image processing. This approach informs the structural improvements made to the GAN generator in the current study, ensuring that essential details are preserved during the fusion process.

Attention mechanisms have also played a significant role in advancing image processing techniques. Zhu et al. [17] introduced the Attention-Unet model, which enhances segmentation accuracy by focusing on relevant features. The incorporation of the Convolutional Block Attention Module (CBAM) in this work builds on similar principles, improving the model's focus on key areas during image fusion. Ma et al. [18] explored the application of Conditional GANs (CGANs) in single image defogging, which aligns with the objectives of enhancing image clarity and detail retention in this study. Their research provided insights into conditional generative frameworks, which have been adapted to improve the performance of the proposed image fusion algorithm.

In addition to these GAN-focused studies, the research by Zhan et al. [19] on feature extraction techniques in medical imaging systems using deep learning offers relevant insights into optimizing models for image fusion tasks. Their work underscores the importance of advanced feature extraction methods in improving image quality and model performance. Further, the study by Yan et al. [20] on neural networks for survival prediction across diverse cancer types highlights the broader applicability of deep learning models in complex scenarios. This research provides a background understanding of how neural networks can be optimized for specific tasks, which is relevant to the optimization strategies used in this paper. Finally, Mei et al. [21] discussed efficiency optimization strategies for large-scale language models, which, while focused on natural language processing, offer valuable insights into general model optimization techniques. These strategies are pertinent to the development of lightweight and efficient GAN models suitable for real-time applications. Xiao et al. [22] also contributed to the discussion on convolutional neural networks (CNNs) by examining their application in cancer cytopathology image classification. While their work is focused on medical imaging, the principles of CNN optimization are relevant to improving the convolutional layers in the proposed GAN model. Together, these studies form a robust foundation for the development of a lightweight, efficient GAN-based image fusion algorithm, designed to maintain high fusion quality while being suitable for deployment in resource-constrained environments.

## III. ALGORITHM PRINCIPLE

Generative Generative adversarial networks (GANs) employed for visible light and infrared image fusion often face challenges due to their substantial model sizes. These large models can hinder effective deployment, delay real-time inference processes, and consume significant computing resources. To tackle these limitations and facilitate the development of a more manageable, lightweight model, an innovative approach involves substituting the standard convolutional layers within the generator with depth-wise separable convolutional layers. This substitution significantly diminishes the computational burden and reduces the total number of parameters required. Furthermore, the integration of a Convolutional Block Attention Module (CBAM) after the dense connection block substantially enhances the image fusion capabilities of the network. By adopting this modified structure, the revised network not only lessens the computational demands and parameter volume but also maintains the high quality of the resultant fused image. This optimized generator architecture, contributing to both efficiency and performance, is depicted in Figure 1.

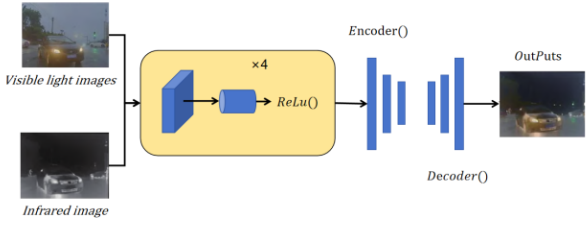

Figure 1 generator structure

The core idea of depth-separable convolution of infrared images is to decompose the complete convolution operation into two steps: depth convolution (Depthwise Convolution, DConv) and point convolution (Pointwise Convolution, PConv). Its structure is shown in Figure 2. This decomposition can significantly reduce the number of parameters, thereby making the network lightweight.

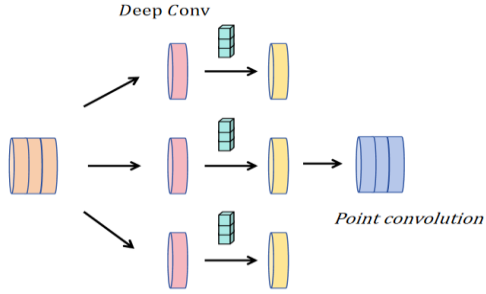

Figure 2 Depth-separable Convolution

Depthwise separable convolution can effectively reduce the number of parameters and computational load. Since the trained generative adversarial network only requires the generator structure for image fusion, replacing conventional convolution with depthwise separable convolution in the generator structure achieves a lightweight algorithm. A more complex network structure is used in the improved generator structure, including multiple depth-separable convolution layers and CBAM attention modules, so that the model can generate more refined fused images.

To enable the network to focus on the critical aspects of the input features and ensure the quality of fused images when dealing with complex and dynamic image fusion tasks, it is essential to consider the relationships between different channels and spatial positions within the source image. This attention to both channel and spatial interactions helps the network more effectively capture and preserve important information during the fusion process. Therefore, a CBAM hybrid attention module is inserted after the densely connected block of the generator.

The CBAM module consists of two modules connected in series. Based on the input feature $F$, it fuses the attention weights in the channel and space dimensions in turn to obtain the refined feature $F''$. The channel attention module is designed to emphasize the channel-specific information within the image, while the spatial attention module focuses on capturing the spatial relationships across the image. The specific calculation process for the channel attention module can be expressed mathematically as follows:

$$M_c(F) = \sigma(MLP(AvgPool(F)) + MLP(MaxPool(F)))$$
$$= \sigma(W_1(W_0(F_{avg}^c)) + W_1(W_0(F_{max}^c)))$$

Among them, AvgPool and MaxPool correspond to average pooling and max pooling operations, respectively. $W_0$ and $W_1$ represent convolution operations; $F_{avg}^c$ and $F_{max}^c$ represent average pooling features and maximum pooling features respectively.

The spatial attention module sequentially applies max pooling and average pooling across the channels of each feature point generated during the convolution process. This results in the extraction of the maximum and average values for each spatial position. These output tensors are then concatenated, and a convolution layer combined with an activation function is used to learn the weights for different spatial positions. This process ultimately generates features with heightened spatial significance. The specific calculation method for the spatial attention module is detailed in the following formula.

$$M_s(F') = \sigma(f^{7 \times 7}([AvgPool(F'); MaxPool(F')]))$$
$$= \sigma(f^{7 \times 7}([F_{avg}^s; F_{max}^s]))$$

As a result, the generator in the lightweight algorithm is more effective at capturing both channel and spatial information within the input feature map. This enhancement improves the model's ability to represent image features and overall performance, leading to the production of higher-quality fused images.

IV. EXPERIMENT

A. Experimental setup

During model testing, the learning rate is initially set to 0.01 and gradually increased to 0.1. The weight decay coefficient is configured at 0.0005, with a batch size of 32. The optimization process employs the adaptive moment estimation (Adam) optimizer with weight decay. All input images are resized to 320×320 pixels, and the model undergoes 500 training epochs.

B. Datasets

To enhance the reliability of the experimental results, the M3FD dataset—an open and calibrated dataset containing both infrared and visible light images—was selected. Established by the School of Software at Dalian University of Technology in 2022, the M3FD dataset comprises 4,200 pairs of infrared and visible light images, representing 4 typical categories and 10 sub-scenes. Compared to traditional datasets such as the TNO and RoadScene datasets, M3FD offers a greater number of registered image pairs. These pairs include infrared and visible light images captured on urban roads under various weather conditions and environmental settings. Additionally, the dataset provides annotations for 33,603 objects across 6 categories, making it particularly well-suited for research on infrared and visible light image fusion and detection algorithms within urban environments.

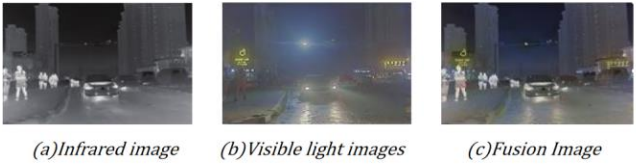

(a) Infrared image    (b) Visible light images    (c) Fusion Image

### C. Experimental Results

In the experiment, TarDAL was used as the benchmark algorithm, and four groups of experiments were conducted, including the benchmark model, the benchmark model combined with the deep separable convolution (TarDAL+DSConv), the benchmark model combined with the CBAM attention mechanism (TarDAL+CBAM), and the lightweight algorithm. It was also compared horizontally with four similar algorithms based on generative adversarial networks for infrared and visible light image fusion, namely DenseFuse, GANMC, U2Fusion, and DDcGAN.

Table 1 Experimental Results

| Model | EN | MI | SF | AG | PNSR | SSIM |
|---|---|---|---|---|---|---|
| DF | 6.338 | 2.690 | 6.360 | 2.127 | 65.979 | 0.734 |
| GAN | 6.655 | 2.658 | 6.067 | 2.074 | 64.552 | 0.739 |
| U2F | 6.601 | 2.673 | 7.534 | 3.215 | 65.457 | 0.720 |
| DDG | 7.420 | 2.445 | 13.767 | 5.494 | 61.328 | 0.517 |
| TarDAL | 7.077 | 4.539 | 10.422 | 3.198 | 63.884 | 0.854 |
| TD+CB | 7.065 | 4.040 | 9.450 | 3.008 | 64.046 | 0.854 |
| TD+DS | 7.046 | 3.866 | 9.458 | 3.060 | 64.140 | 0.852 |
| ours | 7.111 | 4.687 | 9.771 | 3.111 | 63.821 | 0.858 |

The above 8 image fusion algorithms are evaluated according to the above 6 image evaluation indicators, and their average performance is shown in Table 2. According to the results in the table, different algorithms show differences under objective quantitative analysis. Compared with the 4 similar image fusion algorithms, the lightweight algorithm performs better in MI and SSIM, but other indicators are insufficient. Compared with TarDAL, the lightweight algorithm adds CBAM attention module and replaces ordinary convolution with deep separable convolution, which improves EN, MI and SSIM. EN increases by 0.48%, indicating that the image information content increases, and the details and complexity in the image are improved; The mutual information (MI) increases by 3.26%, suggesting that the similarity between the fused image and the source images has improved. This indicates that the fused image preserves a greater amount of valuable information from the original source images; SSIM increases by 0.47%, indicating that the visual quality and structural fidelity of the image are improved, and it is closer to the original image. Compared with TarDAL, the lightweight algorithm has a decrease in SF, AG, and PSNR. SF decreases by 6.25%, indicating that the texture of the image becomes smoother and the local changes of details are reduced; AG decreases by 2.72%, indicating that the image details and edge contrast are reduced; PSNR decreases by 1.0%, indicating that the image quality is reduced to some extent.

Table 2 Comparison of inference delay and parameter quantity of six different algorithms

| Model | Reasoning delay/s | Parameter quantity/10^6 |
|---|---|---|
| DenseFuse | 0.018 | 0.299 |
| GANMcC | 35.093 | 2.276 |
| U2Fusion | 4.244 | 0.659 |
| DDcGAN | 10.476 | 5.553 |
| TarDaL | 0.189 | 1.185 |
| ours | 0.178 | 0.163 |

The comparison of inference delay and parameter quantity of 6 different algorithms is shown in Table 3. The network model structure is tested and the average value of model inference delay of 5 training processes is taken. After comparison, it is found that compared with similar image fusion algorithms, the lightweight algorithm has a lower inference delay and fewer parameters. Compared with TarDAL, the inference delay time of the lightweight image fusion algorithm is slightly shortened, and the model parameter quantity is reduced by 86.24%.

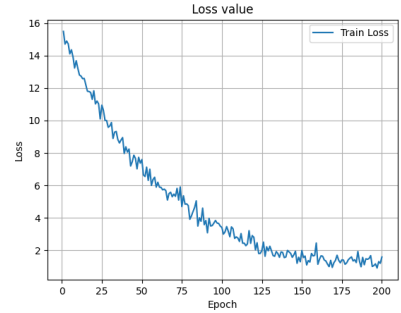

(a) SGD optimizer loss function decline graph

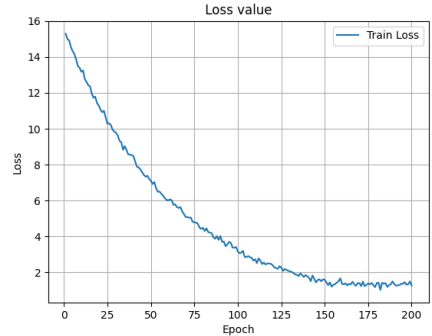

(b) AdamW optimizer loss function decline graph

From the comparison of the loss reduction between SGD and AdamW, the reason why Adam is usually better than SGD is that AdamW has a momentum mechanism. Adam combines the idea of momentum, which helps to accelerate convergence and avoid local minima, further improving training efficiency.

## V. CONCLUSION

This paper presents a groundbreaking lightweight image fusion algorithm that enhances the fusion of visible and infrared images by innovatively modifying the generator within a generative adversarial network (GAN) framework. The integration of the Convolutional Block Attention Module (CBAM) and the implementation of Depthwise Separable Convolution (DSConv) within the generator architecture are key contributions of this work. These enhancements address the common issues of information loss and computational inefficiency in traditional GAN-based fusion methods. By focusing on improving the model's ability to prioritize critical image features, the algorithm ensures that essential details from both visible and infrared images are effectively merged. The use of CBAM allows the model to selectively emphasize significant features, while DSConv reduces the computational complexity without compromising the quality of the fused images. Extensive experiments conducted on the M3FD dataset demonstrate the superior performance of the proposed method in terms of fusion quality, parameter efficiency, and inference speed when compared to existing state-of-the-art algorithms. The results confirm that the proposed algorithm not only produces higher quality fused images but also significantly reduces the computational burden, making it an ideal solution for real-time applications on embedded devices where resource constraints are a critical concern. Furthermore, the successful reduction in model size and inference delay highlights its potential for broader applications in computer vision tasks, such as target detection and scene perception, particularly in environments where rapid processing and deployment efficiency are paramount. Through this research, the potential for lightweight GAN-based models in advancing the field of image fusion and their applicability in diverse, resource-constrained settings has been firmly established, opening new avenues for further exploration and optimization in future studies.

REFERENCES

[1] Liu, Y., Yang, H., & Wu, C. (2023). Unveiling patterns: A study on semi-supervised classification of strip surface defects. IEEE Access, 11, 119933-119946.

[2] Li, P., Lin, Y., & Schultz-Fellenz, E. (2018). Contextual hourglass network for semantic segmentation of high resolution aerial imagery. arXiv preprint arXiv:1810.12813.

[3] Cao, J., Xu, R., Lin, X., Qin, F., Peng, Y., & Shao, Y. (2023). Adaptive receptive field U-shaped temporal convolutional network for vulgar action segmentation. Neural Computing and Applications, 35(13), 9593-9606.

[4] Liu, S., Yan, K., Qin, F., Wang, C., Ge, R., Zhang, K., ... & Cao, J. (2024). Infrared Image Super-Resolution via Lightweight Information Split Network. arXiv preprint arXiv:2405.10561.

[5] Yao, J., Li, C., Sun, K., Cai, Y., Li, H., Ouyang, W., & Li, H. (2023, October). Ndc-scene: Boost monocular 3d semantic scene completion in normalized device coordinates space. In 2023 IEEE/CVF International Conference on Computer Vision (ICCV) (pp. 9421-9431). IEEE Computer Society.

[6] Liu, Z., Wu, M., Peng, B., Liu, Y., Peng, Q., & Zou, C. (2023, July). Calibration Learning for Few-shot Novel Product Description. In Proceedings of the 46th International ACM SIGIR Conference on Research and Development in Information Retrieval (pp. 1864-1868).

[7] Li, S., Kou, P., Ma, M., Yang, H., Huang, S., & Yang, Z. (2024). Application of Semi-supervised Learning in Image Classification: Research on Fusion of Labeled and Unlabeled Data. IEEE Access.

[8] Liu, H., Li, I., Liang, Y., Sun, D., Yang, Y., & Yang, H. (2024). Research on Deep Learning Model of Feature Extraction Based on Convolutional Neural Network. arXiv preprint arXiv:2406.08837.

[9] Creswell, A., White, T., Dumoulin, V., Arulkumaran, K., Sengupta, B., & Bharath, A. A. (2018). Generative adversarial networks: An overview. IEEE signal processing magazine, 35(1), 53-65.

[10] Song, J., & Liu, Z. (2021, November). Comparison of Norm-Based Feature Selection Methods on Biological Omics Data. In Proceedings of the 5th International Conference on Advances in Image Processing (pp. 109-112).

[11] Panfeng Li, Qikai Yang, Xieming Geng, Wenjing Zhou, Zhicheng Ding, & Yi Nian (2024). Exploring Diverse Methods in Visual Question Answering. arXiv preprint arXiv:2404.13565.

[12] Y. Zhong, Y. Wei, Y. Liang, X. Liu, R. Ji, and Y. Cang, "A comparative study of generative adversarial networks for image recognition algorithms based on deep learning and traditional methods," arXiv preprint arXiv:2408.03568, 2024.

[13] Y. Feng, B. Zhang, L. Xiao, Y. Yang, T. Gegen, and Z. Chen, "Enhancing medical imaging with GANs synthesizing realistic images from limited data," in 2024 IEEE 4th International Conference on Electronic Technology, Communication and Information (ICETCI), 2024, pp. 1192-1197.

[14] X. Liu, H. Qiu, M. Li, Z. Yu, Y. Yang, and Y. Yan, "Application of multimodal fusion deep learning model in disease recognition," arXiv preprint arXiv:2406.18546, 2024.

[15] Y. Hu, H. Yang, T. Xu, S. He, J. Yuan, and H. Deng, "Exploration of multi-scale image fusion systems in intelligent medical image analysis," arXiv preprint arXiv:2406.18548, 2024.

[16] Z. Gao, Q. Wang, T. Mei, X. Cheng, Y. Zi, and H. Yang, "An enhanced encoder-decoder network architecture for reducing information loss in image semantic segmentation," arXiv preprint arXiv:2406.01605, 2024.

[17] Z. Zhu, Y. Yan, R. Xu, Y. Zi, and J. Wang, "Attention-Unet: A deep learning approach for fast and accurate segmentation in medical imaging," Journal of Computer Science and Software Applications, vol. 2, no. 4, pp. 24-31, 2022.

[18] R. Q. Ma, X. R. Shen, and S. J. Zhang, "Single image defogging algorithm based on conditional generative adversarial network," Mathematical Problems in Engineering, vol. 2020, no. 1, p. 7938060, 2020.

[19] Q. Zhan, D. Sun, E. Gao, Y. Ma, Y. Liang, and H. Yang, "Advancements in feature extraction recognition of medical imaging systems through deep learning technique," arXiv preprint arXiv:2406.18549, 2024.

[20] X. Yan, W. Wang, M. Xiao, Y. Li, and M. Gao, "Survival prediction across diverse cancer types using neural networks," in Proceedings of the 2024 7th International Conference on Machine Vision and Applications, 2024, pp. 134-138.

[21] T. Mei, Y. Zi, X. Cheng, Z. Gao, Q. Wang, and H. Yang, "Efficiency optimization of large-scale language models based on deep learning in natural language processing tasks," arXiv preprint arXiv:2405.11704, 2024.

[22] M. Xiao, Y. Li, X. Yan, M. Gao, and W. Wang, "Convolutional neural network classification of cancer cytopathology images: taking breast cancer as an example," in Proceedings of the 2024 7th International Conference on Machine Vision and Applications, 2024, pp. 145-149.